\documentclass[fleqn,usenatbib]{mnras}

\usepackage[T1]{fontenc}
\usepackage{ae,aecompl}

\usepackage{epsf,palatino,url,wrapfig,array,setspace,amsmath,amssymb,fancyhdr,multirow,lscape,appendix,rotating}
\usepackage{graphics,epsfig,txfonts}
\usepackage{graphicx}

\usepackage{longtable}
\usepackage{amssymb}
\usepackage{xcolor}
\usepackage{color}
\usepackage{lipsum}
\usepackage{textcomp}
\usepackage{threeparttable}
\usepackage{enumerate}
\usepackage{placeins}
\usepackage{float}
\usepackage[normalem]{ulem}
\usepackage{soul}
\usepackage{blindtext}

\usepackage{tikz,hyperref}
\definecolor{lime}{HTML}{A6CE39}
\DeclareRobustCommand{\orcidicon}{%
	\begin{tikzpicture}
	\draw[lime, fill=lime] (0,0) 
	circle [radius=0.16] 
	node[white] {{\fontfamily{qag}\selectfont \tiny ID}};
	\draw[white, fill=white] (-0.0625,0.095) 
	circle [radius=0.007];
	\end{tikzpicture}
	\hspace{-2mm}
}

\foreach \x in {A, ..., Z}{%
	\expandafter\xdef\csname orcid\x\endcsname{\noexpand\href{https://orcid.org/\csname orcidauthor\x\endcsname}{\noexpand\orcidicon}}
}

\newcommand{\orcid}[1]{\href{https://orcid.org/#1}{\textcolor[HTML]{A6CE39}{\orcidicon}}}

\newcommand{\oergcm}[1]{$10^{#1}$ erg cm$^{-2}$ s$^{-1}$}
\newcommand{\ergs}[1]{$\times 10^{#1}$ erg s$^{-1}$}
\newcommand{\oergs}[1]{$10^{#1}$ erg s$^{-1}$}
\newcommand{\hcm}[1]{$\times 10^{#1}$ cm$^{-2}$}

\newcommand{\ltsima}{$\buildrel < \over \sim$}
\newcommand{\lsim}{\lower.5ex\hbox{\ltsima}}
\newcommand{\gtsima}{$\buildrel > \over \sim$}
\newcommand{\gsim}{\lower.5ex\hbox{\gtsima}}

\newcommand{\swift}{{\it Swift}\xspace}

\newcommand{\xmm}{{\it XMM-Newton}\xspace}
\newcommand{\ROSAT}{\hbox{ROSAT}\xspace}

\newcommand{\rxh}{\hbox{1RXS\,J050526.3-684628}\xspace}
\newcommand{\rxhs}{\hbox{J050526}\xspace}

\newcommand{\gv}{\textcolor{blue}}

\graphicspath{{./}{figures/}{plots/}}

\title[A Slow Evolving Post-Nova SSS in the LMC]{Discovery of a $\sim$30-Year-Duration Post-Nova Pulsating Supersoft Source in the Large Magellanic Cloud}

\author[G. Vasilopoulos et al.]
{G.~Vasilopoulos\orcid{0000-0003-3902-3915},$^1$\thanks{E-mail: georgios.vasilopoulos@yale.edu}
F.~Koliopanos,$^2$
T.~E.~Woods\orcid{0000-0003-1428-5775},$^3$
F.~Haberl\orcid{0000-0002-0107-5237},$^4$
\newauthor
M.~D.~Soraisam,$^{5,6}$
A.~Udalski\orcid{0000-0001-5207-5619}$^7$
\\
$^1$Department of Astronomy, Yale University, PO Box 208101, New Haven, CT 06520-8101, USA \\
$^2$Universit{\'e} de Toulouse; UPS-OMP; IRAP, 31058 Toulouse, France\\
$^3$National Research Council of Canada, Herzberg Astronomy \& Astrophysics Research Centre, 5071 West Saanich Road, Victoria, BC V9E 2E7, Canada\\
$^4$Max-Planck-Institut f\"ur extraterrestrische Physik,Giessenbachstra{\ss}e, 85748 Garching, Germany\\
$^5$National Center for Supercomputing Applications, University of Illinois at Urbana-Champaign, Urbana, IL 61801, USA\\
$^6$Department of Astronomy, University of Illinois at Urbana-Champaign, Urbana, IL 61801, USA\\
$^7$Astronomical Observatory, University of Warsaw, Al. Ujazdowskie 4, 00-478 Warszawa, Poland
}

\date{Accepted XXX. Received YYY; in original form ZZZ}

\pubyear{2020}

\begin{document}
\label{firstpage}
\pagerange{\pageref{firstpage}--\pageref{lastpage}}
\maketitle

\begin{abstract}

Supersoft X-ray sources (SSS) have been identified as white dwarfs accreting from binary companions and undergoing nuclear-burning of the accreted material on their surface. Although expected to be a relatively numerous population from both binary evolution models and their identification as Type Ia supernova progenitor candidates, given the very soft spectrum of SSSs relatively few are known. Here we report on the X-ray and optical properties of \rxh, a previously unidentified accreting nuclear-burning white dwarf located in the Large Magellanic Cloud (LMC).
\xmm observations enabled us to study its X-ray spectrum and measure for the first time short period oscillations of $\sim$170\,s. 
By analysing newly obtained X-ray data by eROSITA,
together with \swift observations and archival \ROSAT data, we have followed its long-term evolution over the last 3 decades. We identify \rxh as a slowly-evolving post-nova SSS undergoing residual surface nuclear-burning, which finally reached its peak in 2013 and is now declining. 
Though long expected on theoretical grounds, such long-lived residual-burning objects had not yet been found.
By comparison with existing models, we find that the effective temperature and luminosity evolution are consistent with a $\sim$0.7 $M_{\odot}$ carbon-oxygen white dwarf accreting $\sim$10$^{-9}~\rm{M}_{\odot}$/yr.
Our results suggest there may be many more undiscovered SSSs and ``missed'' novae awaiting dedicated deep X-ray searches in the LMC and elsewhere.

\end{abstract}




\section{Introduction}\label{sec:intro}

Super-soft X-ray sources (SSSs) are defined by their approximate black-body spectra with temperatures and luminosities of 20-100 eV and $\ge 10^{35}$ erg s$^{-1}$, respectively \citep{1996LNP...472.....G}. 
Many SSS are now understood to be binary systems wherein a white dwarf (WD) undergoes surface nuclear-burning of matter accreted from a companion star \citep{1997ARA&A..35...69K}. These systems may play a vital role in the origin of i-process elements \citep{Denissenkov17}, provide a unique probe of the warm interstellar medium \citep{WG16}, and are an essential benchmark in understanding the evolution of interacting binaries \citep{Chen14, Chen15}. Perhaps most famously, if an accreting WD can grow to reach the Chandrasekhar mass limit ($\approx1.4M_{\odot}$), it may explode as a type Ia supernova.
Although the total contribution of such objects to the observed type Ia rate remains uncertain \citep[see review][]{2014ARA&A..52..107M}, recent abundance measurements suggest a significant fraction of SNe Ia must originate in near-Chandrasekhar mass explosions \citep{Hitomi}.

The Magellanic Clouds harbour a well studied population of SSS \citep{1996LNP...472.....G}.
Their moderate and well known distances, as well as the low Galactic foreground absorption, make SSSs ideal targets for examining their bolometric luminosities and spectral properties.
Here, we provide the first identification of \rxh 
(hereafter \rxhs) as a very long-lived post-nova SSS based on \xmm observations carried out on February 09, 2013 (obsid: 0693450201), and on  October 19, 2017 (obsid: 0803460101). Originally detected as a soft X-ray source in the LMC during the ROSAT all-sky survey \citep{1999A&A...349..389V}, \rxhs has remained uncharacterized until now, likely due to the low statistics in the previously available data.  
In the following, we report the X-ray spectral and temporal properties of the SSS system observed by \xmm, which confirm the nature of this system as being consistent with an accreting white dwarf (WD) undergoing residual nuclear-burning and short-period pulsations. We also identify a possible optical counterpart from observations made by the Optical Gravitational Lensing Experiment (OGLE). 

\section{Data Analysis}\label{sec:data}

\xmm data were analysed by using the Data Analysis software SAS, version 17.0.0 and most recent calibration files. 
To search for background flares, we defined a background threshold of 8 and 2.5 counts ks$^{-1}$ arcmin$^{-2}$ for the EPIC-pn and EPIC-MOS detectors, respectively. Event extraction was performed using the SAS task \texttt{evselect}, with filtering flags (\texttt{\#XMMEA\_EP \&\& PATTERN<=4} for pn and \texttt{\#XMMEA\_EM \&\& PATTERN<=12} for MOS). SAS tasks \texttt{rmfgen} and \texttt{arfgen} were used to create the redistribution matrix and ancillary file. Finally, we performed barycentric corrections to the event arrival times.

The 2013 \xmm (hereafter XMM13) observation (40 ks starting on MJD 56332.5) was affected by major background flares, thus only the first $\sim$30 ks were used for our analysis. 
Additionally, \rxhs was projected in a CCD gap in the EPIC-pn detector ($\sim$80\% of counts were lost).
For the 2017 (hereafter XMM17) observation, \xmm observed \rxhs for 55ks (MJD 58045.2) while data were not affected by background flares.

The source detection was performed simultaneously on all the images using the SAS task {\tt edetect\_chain}. 
To account for the systematic uncertainties we performed boresight corrections based on the source position of known X-ray sources in the field of \xmm.
The X-ray positions were cross-corrected with those of known AGN \citep{2013ApJ...775...92K}, and a boresight correction was computed as the median of the astrometric offsets. 
This resulted in a localization of \rxhs at 
$\alpha_\text{J2000}=05^\text{h}05^\text{m} 21\fs{}67$ and $\delta_\text{J2000}=-68^\circ 45' 38\farcs0$ (0.03\arcsec, $1\sigma$ statistical uncertainty). However, the positional error is dominated by a systematic uncertainty of $\sim$0.5\arcsec\ \citep[see][]{2013A&A...558A...3S}.

\begin{figure*}
	\resizebox{1.02\hsize}{!}{\includegraphics*[width=0.99\columnwidth]{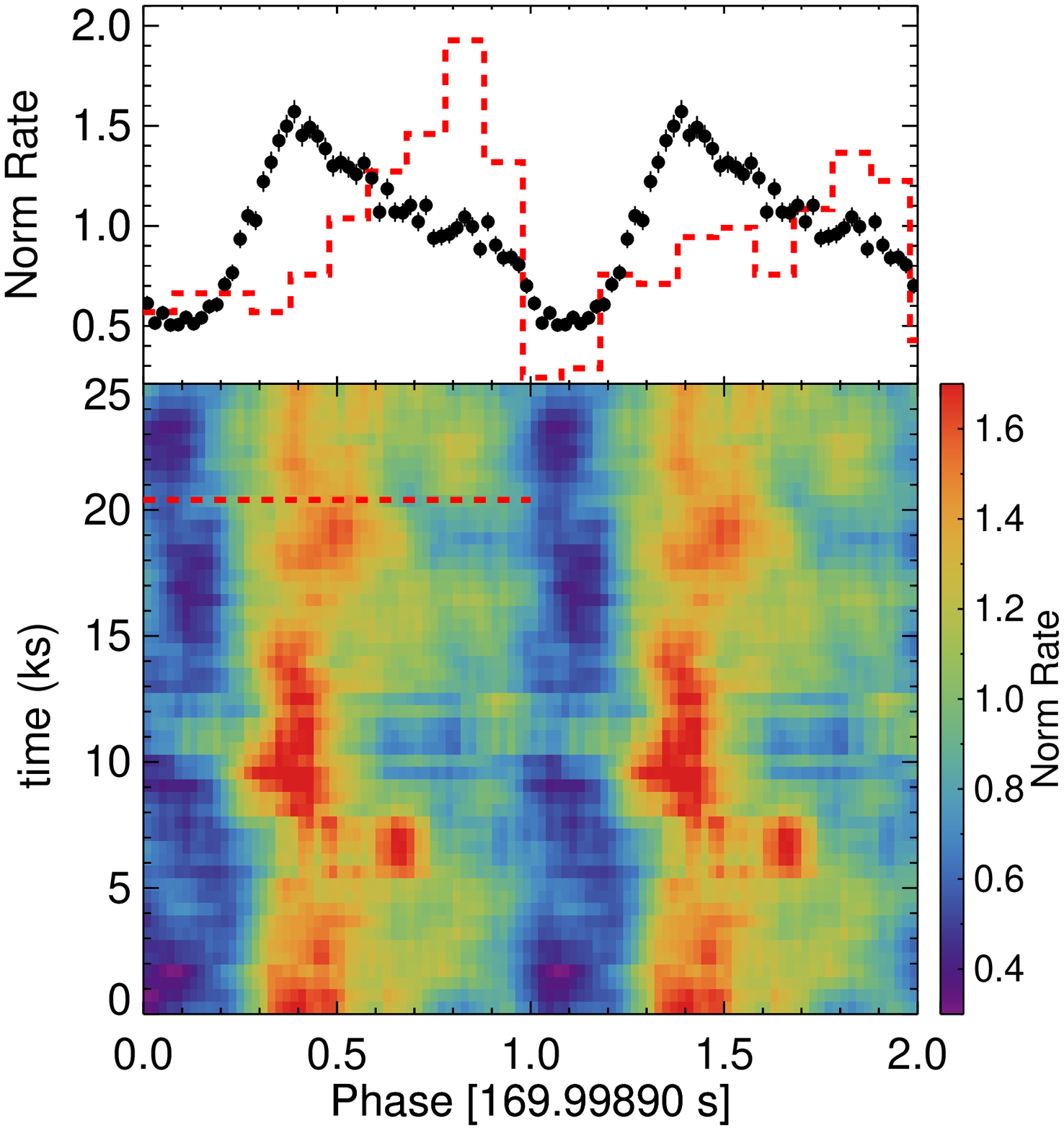}
	\includegraphics*[width=0.99\columnwidth]{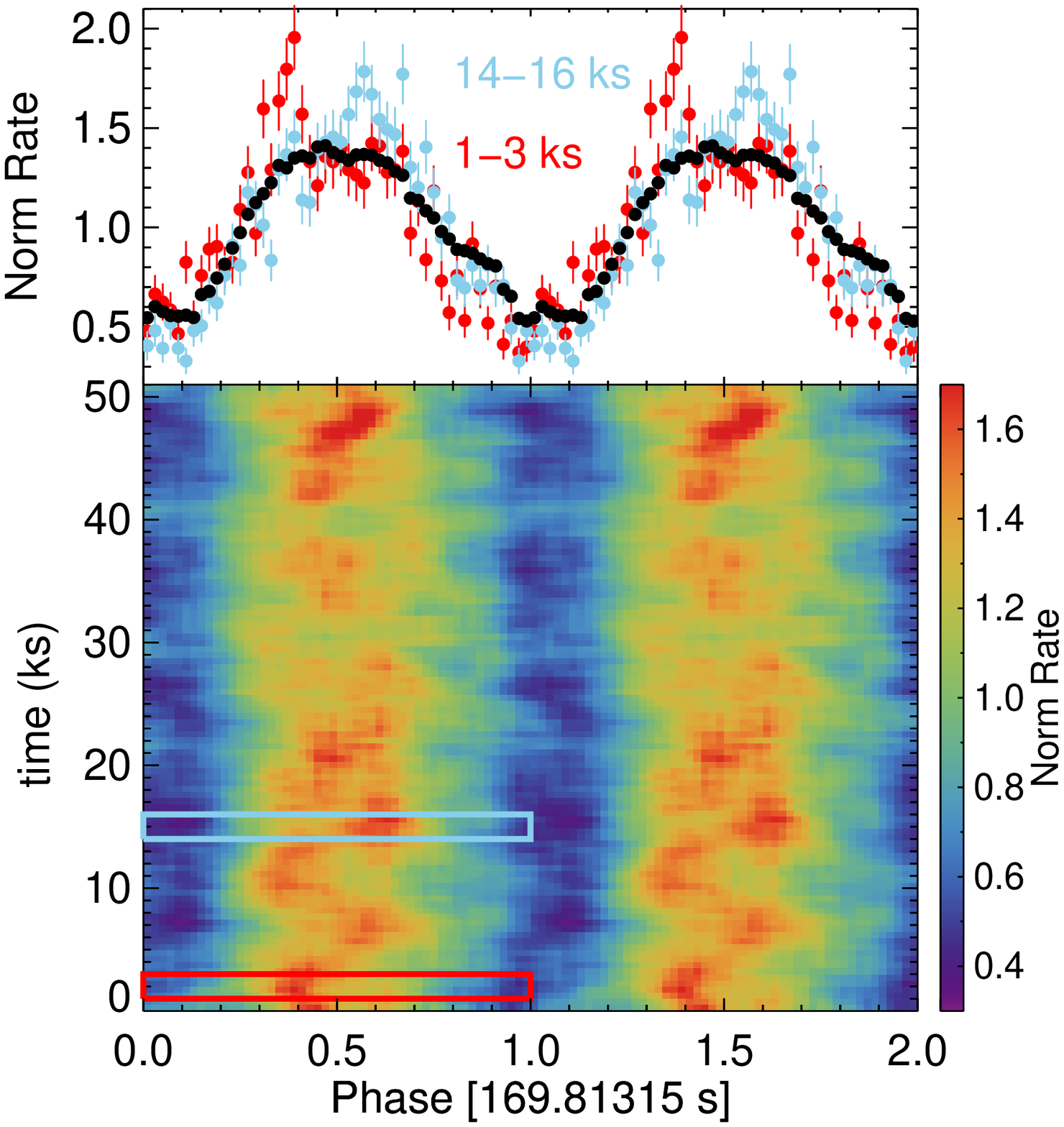}}	
	\vspace{-0.5cm}
    \caption{Time-averaged pulse profile (black points in upper panels) and dynamical pulse profiles (i.e. heat maps in lower panels) of \rxhs for the 2013 (left) and 2017 (right) \xmm data.
    In the heat-map there is evidence for a change in the pulse morphology within each observation. 
    To better demonstrate this variability we created pulse profiles from short intervals. For XMM13 we plot two consecutive pulses on top of the time-averaged profile (dashed red line in left panel).
    For the XMM17 data, we created pulse profiles using 2 ks intervals and over-plotted them over the average profile. Extraction regions for the 2 pulse profiles are marked with coloured boxes on the heat map.  
    }
    \label{fig:PP}
\end{figure*}

\subsection{Timing properties}\label{sec:time}

We searched for a periodic signal in the  barycentric corrected \xmm/EPIC events (merged event lists from the three detectors).
We limited our search to events with detector energies 0.2-1.5 keV.
We used epoch folding implemented through {\tt HENdrics} command-line scripts \citep[][]{2019ApJ...881...39H}.
A period of $\sim$170\,s was detected in all data. 
To estimate period uncertainties we followed \citet[][]{2020A&A...637A..33T,2020MNRAS.494.5350V}. 
We first calculated time of arrivals of individual pulses and then used a Bayesian approach of linear regression to fit them \citep{2007ApJ...665.1489K}.
For XMM13 we found a period $P$ of 170.00$\pm$0.03 s (i.e. $\nu=0.0058824$ Hz), while for XMM17  we found $P=169.813\pm0.014$ s (i.e. $\nu=0.0058888$ Hz). 
This suggests a period derivative of $\dot{P}=-1.26(20)\times10^{-9} s/s$ (i.e. $\dot{\nu}=4.35\times10^{-14}$ Hz/s).

We used the timing solution to create average and dynamical (i.e. heat maps) pulse profiles for the two \xmm observations (see Fig. \ref{fig:PP}). The pulse profiles were created by using all \xmm/EPIC events within the 0.2-1.5 keV energy band, which resulted in 25k and 100k counts from observations XMM13 and XMM17 respectively. The time-averaged pulse profiles are single peaked, however, the dynamical pulse profiles revealed some variability. Specifically, the peak of the pulse modulates between phase 0.4 and 0.7 (see XMM17 pulses in Fig. \ref{fig:PP}). By visually inspecting the light-curve we identified intervals where the profile became double-peaked with a secondary peak at phase 0.8-0.9, this is evident in two consecutive pulses around $\sim$20\,ks into the XMM13 observation (see left panel in Fig. \ref{fig:PP}). 

\subsection{Spectral properties}\label{sec:spec}

All spectra were regrouped to have at least 1 count per bin.
Spectral modeling was performed in {\tt XSPEC} v12.10.1f \citep{1996ASPC..101...17A}, using C-statistics. 
The continuum of SSS spectra can be modeled by either an empirical black-body (BB) model, or by a non-local thermodynamic equilibrium model (NLTE) that provides a more physical description of the WD atmosphere. Both have been successfully used with CCD-quality spectra, where due to the lack of high spectral resolution not all WD atmospheric lines and absorption edges can be resolved \citep[e.g.][]{1994A&A...281L..61G,2001ApJ...550.1007E,2010AN....331..152E,2013A&A...559A..50N}. 
We used publicly available\footnote{\url{http://astro.uni-tuebingen.de/~rauch/TMAF/TMAF.html}} NLTE models for $\log{g}=9$ (in  cgs  units) and pure Hydrogen atmospheres \citep{1999JCoAM.109...65W,2003ASPC..288..103R}.
In the source spectra there is also a high energy tail present, that can be adequately fitted by a power-law (PL) component. From inspection of the X-ray images the hard emission is consistent with a point source and thus is intrinsic to the system. The hard X-ray emission in SSS has been proposed to be due to shocks within the nova ejecta \citep[see case for V1974 Cyg][]{1996ApJ...456..788K}.
To account for the photo-electric absorption we used {\tt tbabs} in {\tt xspec} with Solar abundances set according to \citet{2000ApJ...542..914W} and atomic cross sections from \citet{1996ApJ...465..487V}.
We used two absorption components to account for the Galactic and the intrinsic absorption of the LMC and the source \citep[e.g.][]{2013A&A...558A..74V,2014A&A...567A.129V,2017A&A...598A..69H}.
We fixed the Galactic column density to the value of $6.98\times10^{20}$ cm$^{-2}$ \citep[][]{1990ARA&A..28..215D}.
For the LMC component, elemental abundances were fixed at 0.49 of the solar values \citep{2002A&A...396...53R}, and the column density was set as free fit parameter.

\begin{figure*}
    \resizebox{\hsize}{!}{\includegraphics[clip]{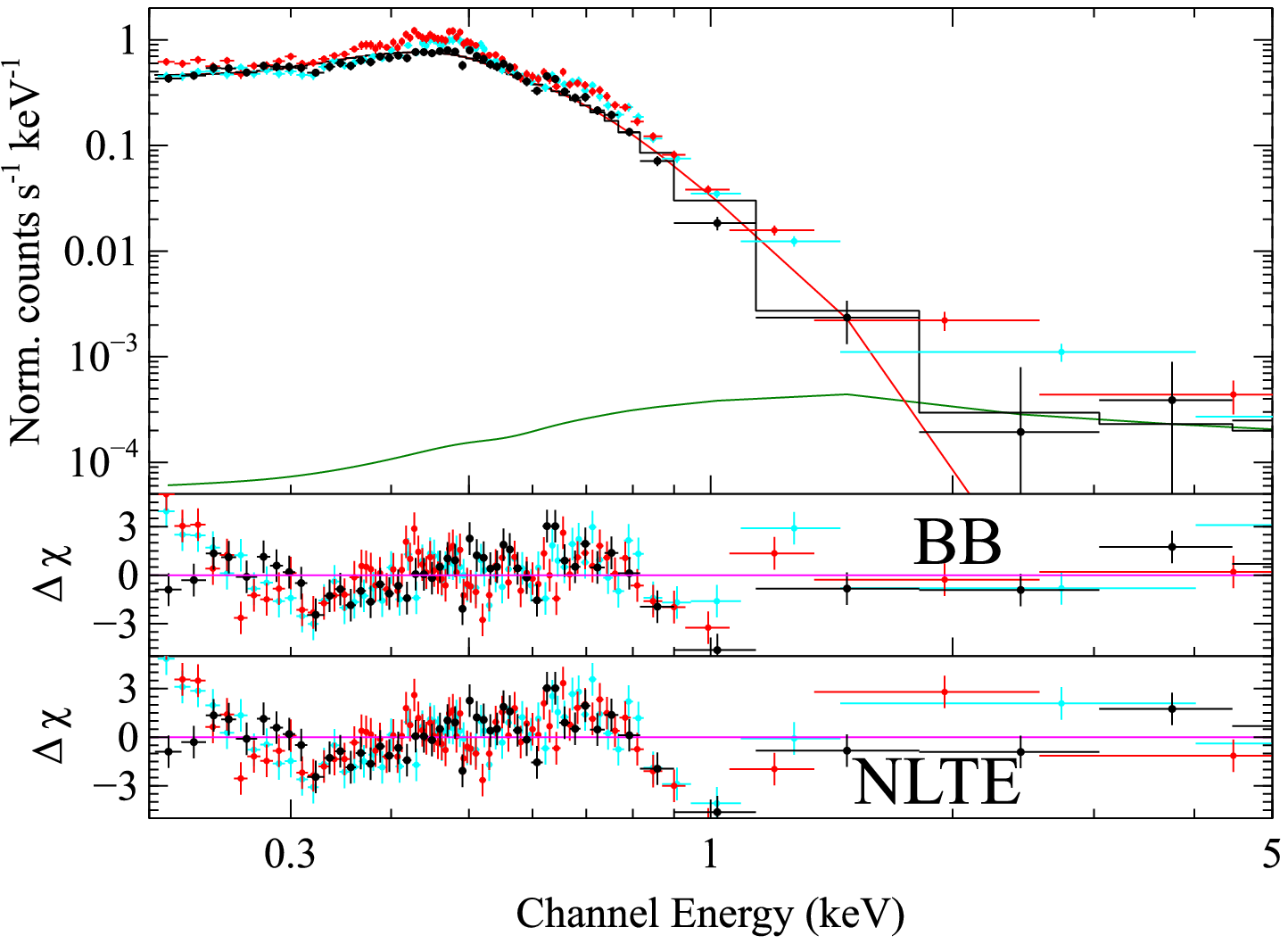}\includegraphics[clip]{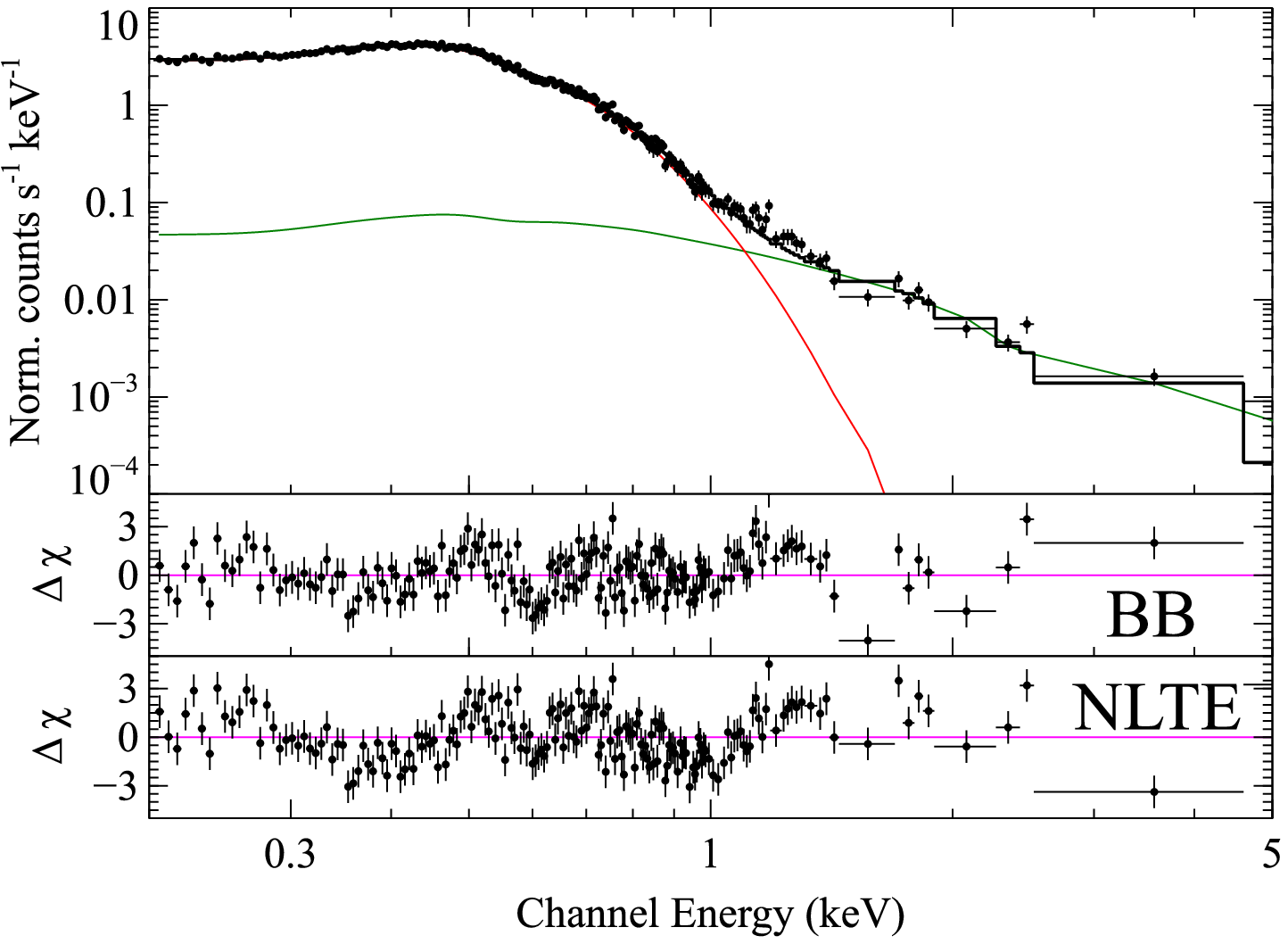}}    
	\vspace{-0.5cm}
    \caption{\emph{Left panel:} 2013 \xmm spectrum of \rxhs. In the upper panel we plot the detector spectra from EPIC-pn (black) and EPIC-mos cameras (red and cyan points). The best fit model (black steps) composed from NLTE (red) and power-law (green) components are also plotted. The lower two panels show the residuals for the models presented in table \ref{tab:spectral}. \emph{Right panels:} same as left but for the XMM13 data and using only the EPIC-pn detector.
    }
    \label{fig:spec}
\end{figure*}

For the XMM2013 data we fitted the model to spectra obtained by all detectors (0.2-10.0 keV band), this was necessary as the source was positioned at the CCD gap of the EPIC-pn detector. For the 2017 fit, we only used the spectra obtained by EPIC-pn (0.2-10.0 keV band), because of the better calibration at lower energies. 
Uncertainties were estimated by a Markov chain Monte Carlo approach and the Goodman-Weare algorithm through {\tt xspec} (Confidence level of 2.706 $\sigma$). For the 2013 data the normalization between the EPIC-mos and pn spectra was different by $\sim$20\%, thus we included this uncertainty in the errors of the reported fluxes. 
The parameters of the best fit are presented in table \ref{tab:spectral} while the spectral fits are shown in Fig. \ref{fig:spec}.

Comparing the BB and NLTE models, the BB shows a better fit, however, this might be expected as NLTE models have been reported to insufficiently model some absorption edges \citep[][]{2010AN....331..152E}. 
Comparing the 2013 and 2017 data the flux (and luminosity) of the system has dropped by a factor of $\sim$2. The emission peak of either spectral component is below the lower limit of the observed spectra. Thus, for either model the fitting is affected by our choice of the spectral range. We used as a lower limit the 0.2 keV range as it is commonly used for SSS systems in the literature, but we note that fitting the data in the 0.3-10.0 keV range generally results in similar spectral shape but larger absorption and larger bolometric $L_x$ (factor of 2-4). 
This could be important when comparing with other systems, where the spectral analysis was performed in different energy bands.

\begin{table*}
\caption{Best-fit parameters of spectral models}\label{tab:spectral}
\begin{threeparttable}[b]
\begin{tabular*}{\textwidth}[t]{p{0.09\textwidth}p{0.1\textwidth}p{0.13\textwidth}p{0.13\textwidth}p{0.13\textwidth}p{0.13\textwidth}p{0.17\textwidth}}
\hline
\hline\noalign{\smallskip} 
& & \multicolumn{2}{c}{XMM - 2013} & \multicolumn{2}{c}{XMM - 2017}  &  \\
\multicolumn{2}{c}{Component Parameters}  & BB model & NLTE model & BB model & NLTE model  & Units  \\
\hline  
\hline\noalign{\smallskip}  
 & $N_{\rm{H}}$ LMC$^{(a)}$  & 2.0$^{+0.7}_{-0.7}$ & 6.7$^{+0.7}_{-0.8}$ & 2.2$^{+0.5}_{-0.4}$& 6.9$^{+0.6}_{-0.5}$ & $10^{20}$cm$^{-2}$\\
{\tt Bbody} & k${\rm T_{BB}}$ & 82.2$^{+1.4}_{-1.1}$ & - & 78.5$^{+1.1}_{-1.1}$& - & eV\\\noalign{\smallskip}
 & $R_{BB}$ $^{(b)}$ & 1060$^{+80}_{-80}$ & - & 830$^{+50}_{-50}$& - &  km\\\noalign{\smallskip} 
{\tt NLTE} & k${\rm T_{WD}}$ & - &360$^{+34}_{-15}$ & -& 328$^{+10}_{-7}$ & $10^3$ Kelvin\\\noalign{\smallskip} 
 & $R_{WD}$ $^{(c)}$& - & 14700$\pm$2900 & -& 13400$\pm$1200 &  km\\\noalign{\smallskip}
{\tt PL} & $\Gamma$ & 1.7$^{+0.6}_{-0.5}$ & 0.3$^{+0.6}_{-0.8}$ & 2.5$^{+0.5}_{-0.4}$& 1.2$^{+0.3}_{-0.3}$ & -\\\noalign{\smallskip}  
 & Norm$^{(d)}$ & 5.8$^{+1.6}_{-0.9}$ &  6.5$^{+2.7}_{-2.6}$  & 5.6$^{+1.8}_{-1.0}$& 4.2$^{+0.7}_{-0.5}$ & \oergs{34}\\\noalign{\smallskip} 
  \hline\noalign{\smallskip}  
$C_{\rm stat}/DOF$  &  & 629.2/433 & 761.5/433 & 294.8/185& 355.9/185 & \\\noalign{\smallskip} 
$F_{X,BB}$ & (0.2-2.0) & 4.1$\pm$0.4 & 4.0$\pm$0.4 & 2.06$\pm$0.1& 2.07$\pm$0.1 & \oergcm{-12} \\\noalign{\smallskip}
$L_{X,BB/WD}$ & (0.2-1.0)$^{(e)}$  & 4.8$^{+0.8}_{-0.7}$ & 11.2$^{+2.3}_{-1.2}$ & 2.4$^{+0.1}_{-0.2}$ & 6.0$^{+0.4}_{-0.5}$ & \oergs{36} \\\noalign{\smallskip}
$L_{X,WD}$ & Bolom.$^{(f)}$ & 6.6$^{+0.5}_{-0.4}$ & 25$^{+4}_{-3}$ & 3.4$^{+0.2}_{-0.3}$ & 14.7$^{+1.9}_{-1.6}$ & \oergs{36} \\\noalign{\smallskip}
  \hline\noalign{\smallskip}  
\end{tabular*}
\tnote{(a)} Column density of the absorption component with LMC abundances, column density of Galactic absorption was fixed to 6.98$\times10^{20}$cm$^{-2}$ (see text for details).
\tnote{(b)} BB radius was estimated from the normalization of the model, assuming a distance of 50 kpc.
\tnote{(c)} The size of the WD can be estimated assuming $L_{\rm X}=4\pi R_{WD}\sigma{T_{WD}^4} $.
\tnote{(d)} Absorption corrected luminosity of the PL component in the 0.3-10.0 keV band. 
\tnote{(e)} Absorption corrected luminosity of the BB/NLTE component in the 0.2-1.0 keV band.
\tnote{(f)} Bolometric $L_{\rm X}$ of the BB/NLTE component.
\end{threeparttable}
\end{table*}

\begin{figure*}
    \resizebox{0.5\hsize}{!}{\includegraphics[angle=0,clip,bb=0 0 640 597]{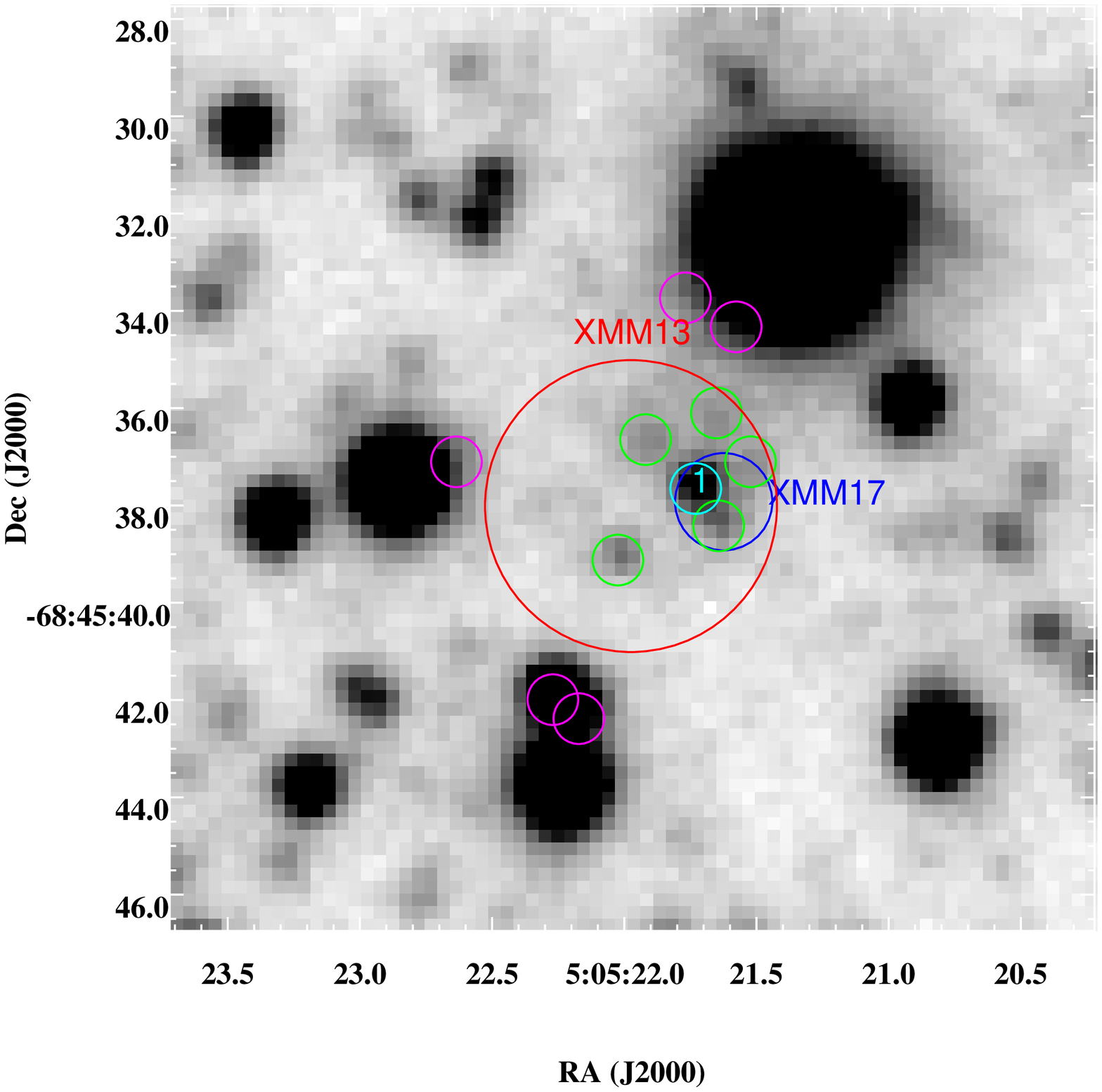}}
    \includegraphics[width=\columnwidth,clip,trim=0 -15 0 0]{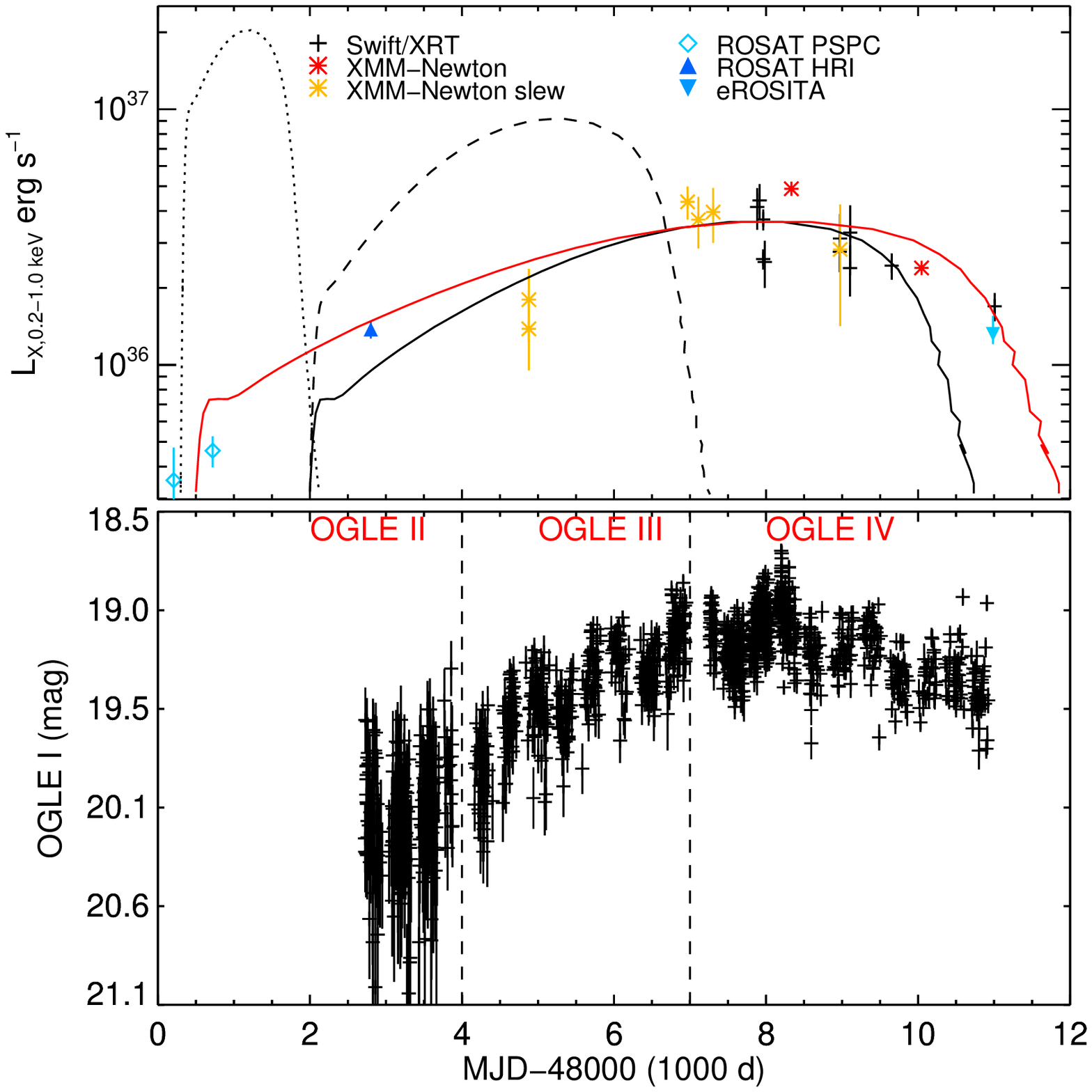}
    \vspace{-0.2cm}
    \caption{\emph{Left:} OGLE finding chart, the red circle with 3\arcsec\ radius is centered on the \xmm 2013 detection, the blue circle with 1\arcsec\ radius marks the location of \rxhs based on the 2017 \xmm detection after boresight correction. The most probable counterpart is located at the center of the image and is marked with a cyan circle. All other counterparts within 4.5\arcsec\ of the \xmm position are marked with various colours, with magenta being the least likely. \emph{Right - Top panel:} X-ray light curve of \rxhs based on all the available X-ray detections. The Y-axis corresponds to the absorption corrected luminosity (0.2-1.0 keV band) of the BB models presented in Table \ref{tab:spectral}. \swift/XRT and \ROSAT count-rates were converted adopting the spectral model parameters derived from the \xmm/EPIC spectra. 
    Overplotted are theoretical predictions of post-nova evolutionary tracks from \citet{2016MNRAS.455..668S}, and for WD masses of 0.9 M$_\odot$ (dotted line), 0.8 M$_\odot$ (dashed line) and 0.7 M$_\odot$ (solid black line), we also plot the 0.7 M$_\odot$ model stretched by 30\% in time (red solid line). \emph{Right - Bottom panel:} OGLE I band light curve of LMC510.12.65281 (cyan circle left panel). Data obtained during the OGLE-II, III \& IV phase.
    }
   \label{fig:ogle}
 \end{figure*}

\subsection{Long term X-ray variability}
We now turn to studying the historic X-ray light-curve of the \rxhs with data obtained over $\sim$30 years from multiple observatories.
The first X-ray detection was made during the \ROSAT all-sky survey in November 1990 \citep[][]{1999A&A...349..389V}. Additional \ROSAT detections followed during pointed PSPC observations on 9 April 1992 \citep[source 715;][]{1999A&AS..139..277H} and HRI observations on 19 December 1997 \citep[source LMC 23;][]{2000A&AS..143..391S}.
Apart from the 2 \xmm pointed observations, \rxhs has been detected 6 times in the \xmm slew survey \citep{2008A&A...480..611S}.
\rxhs was also detected in 11 \swift/XRT pointings between 2011 and 2020 \citep{2014ApJS..210....8E,2020ApJS..247...54E}.
To calculate average count rates for all \swift/XRT detections, we analyzed available data through the \swift science data centre following \citet{2007A&A...469..379E,2009MNRAS.397.1177E}. To convert count rates from all other instruments to unabsorbed $L_{X}$ in the 0.2-1.0 keV band we adopted the best fit BB model from XMM17 (see Table \ref{tab:spectral}).
At this point we comment on adopting a constant k${\rm T_{BB}}$ for the WD pseudophotosphere to convert count rates to unabsorbed $L_{\rm X}$. 
Unfortunately, the lack of observations with high statistics during the early states does not allow us to perform spectral fit to the data.
However, it is expected from theory \citep[see, e.g.,][]{2013ApJ...777..136W} that k${\rm T_{BB}}$ evolves during the post nova phase. For low mass WD, k${\rm T_{BB}}$ can change by a factor 1.5 over 10s of years \citep{2016MNRAS.455..668S}, which would result in an overestimate of $L_{\rm X}$ when assuming constant k${\rm T_{BB}}$ at earlier times. However, this could be compensated by an increased column density in earlier times, as has been suggested by observations of the initial evolution of other post novae \citep{2010MNRAS.401..121P} such that we get a similar unabsorbed $L_{\rm X}$ for a slightly reduced k${\rm T_{BB}}$ with a bit higher column density.

During the course of the first all-sky survey (eRASS1) \rxhs was monitored in May 2020 by the eROSITA instrument on board the Russian/German Spektrum-Roentgen-Gamma (SRG) mission \citep{2012arXiv1209.3114M}. 
Between MJD\,58981.37 and 58989.37, \rxhs was scanned 49 times accumulating a total exposure time of $\sim$1667\,s.
We extracted a combined spectrum from the five eROSITA CCD cameras with on-chip Al blocking filter. The other two cameras suffer from optical light leakage which requires more complicated calibration, before they can be used for reliable spectral analysis. We fitted the spectrum with the same black-body model as used for the \xmm spectra. The best-fit parameters were determined to $N_{\rm{H}}$ LMC = 0 (upper limit 1.8\hcm{20}) and 
k${\rm T_{BB}}$ = 85.6 $\pm$ 4 eV, which results in $L_x = 1.32^{+0.23}_{-0.11}$\ergs{36} (0.2-1.0 keV).

In Fig. \ref{fig:ogle} we present the resulting X-ray light-curve based on all available X-ray data. As we will discuss in $\S$\ref{sec:postnova} this will enable comparison with available theoretical models.

\subsection{Possible optical counterpart}

We searched the available catalogues for possible optical counterparts.
However, many optical surveys could not deliver the desired resolution and sensitivity \citep[e.g.][]{2002ApJS..141...81M,2004AJ....128.1606Z}. Nevertheless, the region of interest was studied by the OGLE survey and several stars were detected within the X-ray error circle \citep{2000AcA....50..307U}.

For the field around \rxhs, OGLE provides more than 20 years of monitoring data.
OGLE images are taken in the V and I filter pass-bands (B filter was also used during  OGLE phase II), while photometric magnitudes are calibrated to the standard VI system \citep{2015AcA....65....1U}. 
There are a few possible optical counterparts located near the X-ray position (Fig. \ref{fig:ogle}). For completeness, we extracted the optical light-curves of all 11 systems located within 4.5\arcsec\ of the uncorrected X-ray position. We investigated all the extracted I band light-curves, and noted that among them only one showed evidence of significant variability (OGLE ID: LMC510.12.65281). The long-term optical variability seems to correlate well with the long-term evolution of the X-ray luminosity of the system (right panel of Fig. \ref{fig:ogle}).
In close binary SSSs, the optical light is dominated by the illuminated low-mass donor star and the accretion disk around the WD \citep{1996LNP...472.....G}, thus the optical light-curve provides strong evidence that this is the correct counterpart of \rxhs. 
The coordinates of the proposed OGLE counterpart are R.A. = 05$^{\rm h}$05$^{\rm m}$21\fs79 and Dec. = --68\degr45\arcmin37\farcs9 (J2000). 
The OGLE II photometric data obtained during the lowest flux state of the optical counterpart can provide an upper limit for the mass of the proposed counterpart \citep{2000AcA....50..307U}. 
Photometric values can be corrected for reddening, by adopting a Galactic extinction curve \citep{1999PASP..111...63F}.
By using an $E(B-V)$ extinction value of 0.055 \citep[][]{2020arXiv200602448S} and the LMC distance module of 18.476 mag \citep[][]{2019Natur.567..200P} we find absolute magnitudes of B$\sim$1.627, V$\sim$1.53 and I$\sim$1.47 (average during OGLE II phase); if we assume that the optical light originates exclusively from the star (neglecting both the accretion disk and irradiation), this is consistent with a main sequence star close to $2M_{\odot}$. We return to this point in $\S$\ref{sec:postnova}.

Given the binary nature of the system it is possible that a periodical signal due to the orbital motion is imprinted in the optical lightcurve. To test this we computed the Lomb-Scargle periodogram \citep{2018ApJS..236...16V} of the complete OGLE data-set. for completeness, we also limited our search to one year long chunks of data. We focused on periods between 0.01 d and 100 d. No periodic signal was identified, and the only peaks in the periodogram were consistent with the OGLE window function (1d, 0.5d, 0.33d and so on).

\section{Discussion}

\begin{figure*}
    \resizebox{\hsize}{!}{\includegraphics[angle=0,clip,trim=0 0 0 0]{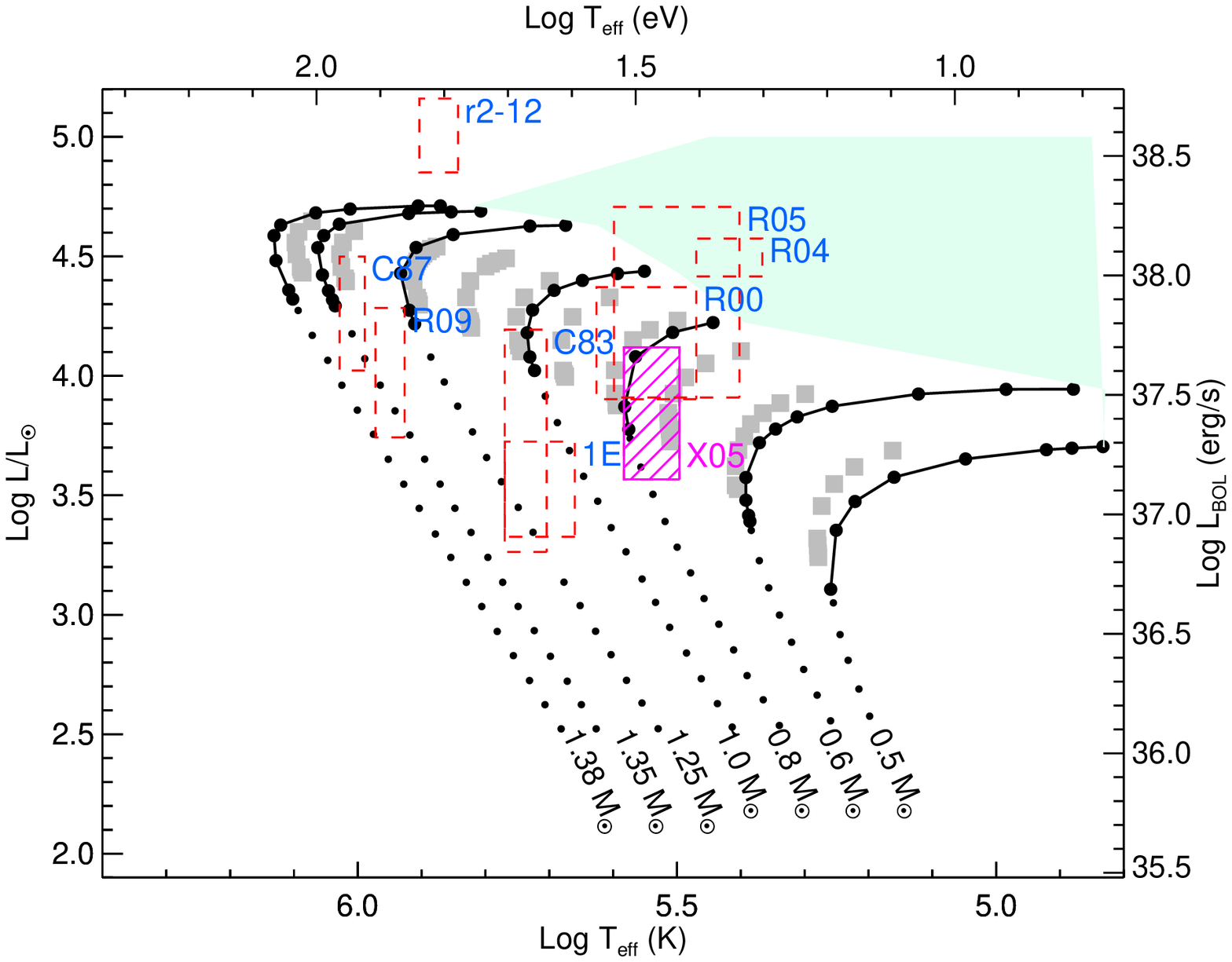}}
        \caption{H-R diagram showing several SSSs; (R09) RXJ0925.7-4758; (C83) CAL 83; (1E) 1E 0035.4-7230; (R00) RX J0019.8+2156; (R04) RX J0439.8-6809; (R05) RX J0513.9-6951 \citep[see][]{2007ApJ...663.1269N,2004ApJ...612L..53S}, (C87) CAL 87 \citep[values corrected for obscuration;][]{2001ApJ...550.1007E}; r2-12 \citep{2008ApJ...676.1218T} and \rxhs (X05) that we present in this paper and the spectral parameters of the NLTE model (we used 5\ergs{37} as an upper limit for $L_x$).
        Also shown are the stable-burning models of \protect\cite{2013ApJ...777..136W} (\gv{gray squares}), and the stable-burning (connected large black dots) and nova (small black dot) white dwarf models of \protect\cite{2007ApJ...663.1269N}. The latter's ``red giant'' or optically-thick wind regime (maximally-accreting) is denoted by the green shaded region. 
        }
   \label{fig:HR}\vskip0.5cm
 \end{figure*}

\subsection{The Nature of J050526}

Examining the evolution of \rxhs over the last 30 years, the source experienced over a ten-fold increase in luminosity between November 1990 and 2011, though initially rising only gradually over the first decade.
In order to constrain the properties of the accreting WD undergoing this eruption, it is illustrative to compare the source radii found from our spectral fitting with that expected from theory. For cold, non-accreting carbon-oxygen WDs, the theoretical mass-radius relation gives \citep{Panei2000}:

\begin{equation}
\frac{R(M)}{R_{\odot}} = 0.0126\left(\frac{\rm{M}}{\rm{M}_{\odot}}\right)^{-1/3}\left(1 - \left(\frac{\rm{M}}{1.456M_{\odot}}\right)^{4/3}\right)^{1/2}
\end{equation}

\noindent Looking first to our BB fits, we find that even at its greatest extent our best-fitting WD radius is consistent only with an extremely massive ($\gtrsim$1.4$M_{\odot}$), extremely compact WD (or perhaps a small region on the WD surface, see further discussion below). This is strongly contradicted, however, by the low (as derived from our black-body fits) luminosity and long timescale of the luminosity evolution observed for \rxhs -- even at peak luminosity, the implied accretion rate in the steady-hydrogen burning regime ($\dot m = L/(\epsilon_{H}X) \approx 3.4\times 10^{-8} M_{\odot}$/yr, with $\epsilon_{H} \approx 6.4e18$ erg/g the energy release due to nuclear burning of hydrogen, and $X \approx 0.72$ the mass fraction of hydrogen) is well below the threshold for steady-burning at this mass \citep[$\sim \rm{few} \times 10^{-7}M_{\odot}$/yr,][]{2007ApJ...663.1269N}, and for such massive WDs at lower accretion rates, post-nova SSSs evolve on timescales measured in days, not years \citep{2013ApJ...777..136W}. We conclude that in this case black-body models are inadequate; detailed NLTE WD atmospheric models are essential in interpreting the soft X-ray spectrum of \rxhs \citep[at least without additional constraints from multiwavelength data, see][for further discussion]{Skopal2015}.

Turning to our NLTE fits (Table \ref{tab:spectral}), we find approximate radius $\sim$15,000\,km;
substantially larger than even the lowest mass non-accreting carbon oxygen WDs. This is consistent (see Fig. \ref{fig:HR}), however, with the inflated photospheres expected in a WD which is undergoing residual nuclear-burning of hydrogen in its remaining envelope after a nova eruption \citep[e.g.,][]{2013ApJ...777..136W}. Alternatively, this could be an indication of a magnetic WD, as such systems exhibit larger radii compared to non-magnetic ones \citep{2000ApJ...530..949S}.

In the upper right panel of Fig. \ref{fig:ogle}, we compare the observed X-ray luminosity evolution as measured with \ROSAT, \swift, and \xmm (0.2-1.0 keV band), with theoretical models of the post-nova hydrogen-burning SSS phase for a 0.7, 0.8, and 0.9 $M_{\odot}$ WD accreting $10^{-9}M_{\odot}$/yr \citep{2013ApJ...777..136W, 2016MNRAS.455..668S}. 
The slow rise and subsequent relatively fast decline from a peak luminosity of $\approx 2.5\times 10^{37}$erg/s closely resembles the predicted evolution of a slowly accreting WD undergoing a post-nova SSS phase.
In particular, we find that the model for a $0.7~M_{\odot}$ WD most closely resembles the observed luminosity evolution of \rxhs,
although the predicted duration of the peak emission  ($L_{x}\gtrsim\times10^{36}~{\rm erg s^{-1}}$) exhibits a somewhat shorter timescale compared to the observed light curve.

Before continuing our comparison with numerical results, we must address two primary uncertainties in the post-nova models, namely, the mass-loss mechanism during the nova outburst, and mixing between the WD core and accreted matter. Wolf et al.\ used two prescriptions in their post-nova MESA models -- super-Eddington wind (SEW), and Roche lobe (RL) overflow. The former is more appropriate for massive WDs ($\gtrsim1~M_{\odot}$), which become super-Eddington early and thereby never expand to their bigger RL radii. \citet{2016MNRAS.455..668S} used only the SEW models to construct the theoretical light curves since bright post-nova SSSs, which arise from massive WDs, are expected to be observed generally. On the other hand, the lower mass WDs tend to fill their RL radii before their luminosity becomes super-Eddington and thus, the RL overflow prescription may be better suited for such WDs. Theoretical light curves for these models are, however, not available. The wind prescription removes more mass than RL overflow, hence the SSS duration is longer in the latter case (by a factor $>5$ for 0.7$M_{\odot}$; see \citealt{2013ApJ...777..136W}). Furthermore, mixing has not been incorporated in the MESA models. Mixing leads to a more violent outburst, which ejects more mass and results in a reduced amount of hydrogen remnant to burn and consequently, shortens the post-nova SSS duration. However, quantifying this effect is difficult (and beyond the scope of this paper). In Fig. \ref{fig:ogle} (upper right panel), the red curve shows the 0.7$M_{\odot}$ model light curve stretched by a factor 1.3, which, interestingly, matches the data well, indicating that a combination of the two uncertainties---mixing and mass-loss---can account for the discrepancy between the observations and model. Thus, all this evidence points to \rxhs as likely a post-nova SSS. We cannot further constrain its nature without additional available models, which we therefore reserve for future work.

\subsection{A post-nova SSS-irradiated donor?}
\label{sec:postnova}

Also requiring further study is the possibility that LMC510.12.65281 is the optical counterpart of \rxhs. If this is the case, what has caused its optical luminosity to rise and fall so closely in tandem with the post-nova SSS X-ray luminosity, long after what would have been the peak in the optical emission of the nova itself? If we naively interpret the emission as arising from the companion alone, its V magnitude and B-V colour are consistent with a $\sim$2$M_{\odot}$ main sequence or early subgiant star, however the strong evolution in its luminosity and its correlation with the soft X-ray flux suggest an additional component. During this time, the inferred WD photospheric radius is too large, and the accretion rate too low, for the disk luminosity to greatly exceed $\sim$$L_{\odot}$. At the same time, the necessarily falling density of the expanding nova ejecta would suggest it is unlikely that the rising optical luminosity could be powered by photoionization of this material by the WD. 

Another possibility is that the residual nuclear-burning luminosity of the WD irradiates the donor, with a fraction of this flux consequently being re-emitted in the optical. Approximating the donor as spherically-symmetric, and assuming it is just filling its Roche lobe, from the vantage point of the WD it will subtend an area on the sky with an angular radius $\theta$, i.e.: 

\begin{equation}
    \tan(\theta) \approx \frac{R_{\rm{donor}}}{a} = \frac{0.49q^{2/3}}{0.6q^{2/3} + \ln(1 + q^{1/3})}
\end{equation}

\noindent where $R_{\rm{donor}}$ is the donor radius, $a$ is the separation between the donor and the WD, and their ratio depends only on the mass ratio $q = M_{\rm{donor}}/M_{\rm{WD}}$ \citep{1983ApJ...268..368E}. For $M_{\rm{WD}} \approx 0.7M_{\odot}$, a donor mass of 1--2$M_{\odot}$ gives q $\approx$ 1.43 -- 2.85, $R_{\rm{donor}}/a \approx 0.41$--0.47, and $\theta \approx$ 0.39--0.44. This means the irradiated donor intercepts as much as $\pi \theta^{2}/4\pi \approx$ 4 -- 5\% of the post-nova SSS's luminosity, only a fraction of which need be re-emitted by the donor's envelope in the I band in order to account for the light curve of LMC510.12.65281. Notably, if we adopt a representative donor radius of $\sim$ 1-- 2 $R_{\odot}$, we may also infer a binary orbital period of $\approx$0.2--0.6 days, comparable to the short term $\sim$0.5 magnitude optical
variations seen throughout the OGLE light curve (recall Fig. \ref{fig:ogle}). Before speculating further, however, follow-up optical spectroscopy will be essential in order to confirm the nature of LMC510.12.65281.

\subsection{Origin of the 170s pulsational period}

The high-frequency pulsations exhibited in the X-ray light curve of \rxhs add a further commonality with most, if not all SSSs \citep[see e.g.,][]{2015A&A...578A..39N}, albeit with a slightly longer period than the 10 -- 100s pulsations typically associated with this class. The origin of SSS pulsations remains a mystery, however the shortest period pulsations are generally argued to be associated with either the rotational period of the WD \citep[e.g., ][]{2014MNRAS.437.2948O} or g-mode oscillations in the nuclear-burning envelope \citep{2003ApJ...584..448D}.

In the rotational period interpretation, the WD has been spun-up by accretion, and accreting matter is funneled by a strong magnetic field from the Keplerian disk toward the WD's poles (the system is an intermediate polar). Such an interpretation has been put forward for the persistent supersoft source r2-12 in M31 \citep{2002ApJ...577..738K}, with a similar pulse period of $\sim$218\,s \citep{2008ApJ...676.1218T}. 
Assuming a Keplerian accretion disk, the torque induced onto the WD due to mass accretion will be maximum when the magnetospheric radius $R_{M}$ is equal to the co-rotation radius (i.e 43000 km).
This is no different than the case of an accreting neutron star, where the torque due to mass accretion is $N_{\rm acc} = \dot{M}\sqrt{GM_{WD}R_{M}}$  \citep[e.g. ][]{2018A&A...620L..12V,2019MNRAS.488.5225V}.
Thus the maximum spin-up rate due to accretion would be: 

\begin{equation}
    \dot P \sim -2.2\times 10^{-15} \rm{ss}^{-1} ~ \dot{M}_{19}I_{\rm{50}}^{-1}m_{\rm{WD}}^{2/3}P^{7/3}
\end{equation}

\noindent where $I_{\rm{50}}$ is the WD's moment of inertia in units of $10^{50}~g~cm^{-2}$, $\dot{M}_{19}$ is the mass accretion rate in units of $10^{19}\rm{g/s}$ and $m_{WD}$ the WD mass in units of $M_{\odot}$.
As a crude approximation, we may take the WD as a constant density sphere within the cool WD radius $R(0.7M_{\odot}) \approx 0.011R_{\odot}$ (recall equation 1), ignoring the much lower density envelope. This gives us $I_{50}\approx3.4$ and an upper bound on $\dot P \sim -5\times 10^{-13}\rm{ss}^{-1}$, about 2000 lower than the observed $\dot P$ for \rxhs.
Assuming that the periodic modulation is indeed due to rotation, an alternative origin for the period's evolution may be that \rxhs hosts a young contracting WD. This scenario has been proposed for other systems, and can produce the observed $\dot{P}$ for a range of WD masses with ages below 1 Myr
\citep[see][]{2018MNRAS.474.2750P}.

Otherwise, this would appear to leave g-mode oscillations in the nuclear-burning envelope as the only viable mechanism to explain the pulsations observed in \rxhs. It should be noted, however, that numerical models which have attempted to simulate such non-radial oscillations, driven by the sensitivity of nuclear-burning to compression (the $\epsilon$-mechanism) at the base of the envelope, have predicted much shorter period oscillations to be excited than are observed in known SSSs \citep{2018ApJ...855..127W}. This problem remains in need of further investigation.

\section{Conclusions}

Largely disregarded upon its initial identification with \ROSAT as a nondescript X-ray source, the $\sim$2013 soft X-ray peak and subsequent decline of \rxhs have revealed it to be a remarkably long-lived post-nova SSS, with a WD mass below that common among other known SSSs but typical of the broader accreting WD population \citep[e.g.,][]{2011A&A...536A..42Z}. Indeed, \rxhs is the longest-duration post-nova SSS yet confirmed as such \citep{2008AJ....135.1328N, 2014A&A...563A...2H}, although amongst the known SSS population of e.g., M31 there are likely many long-lived post-novae awaiting further confirmation from long-term follow-up surveys \citep{2006ApJ...643..844O, 2014A&A...563A...2H, 2016MNRAS.455..668S}. 
As such, \rxhs provides an invaluable probe of the poorly-understood, but likely well-populated, long/soft/moderately faint segment of the parameter space of WD X-ray transients, and a natural laboratory for future X-ray pulsation and irradiation studies.

\section*{Acknowledgements}
The authors would like to thank the anonymous referee for their comments and input that helped improved the manuscript.
This research has made use of data and/or software provided by the High Energy Astrophysics Science Archive Research Center (HEASARC), which is a service of the Astrophysics Science Division at NASA/GSFC.
Based on observations using: \xmm, an ESA Science Mission with instruments and contributions directly funded by ESA Member states and the USA (NASA); \swift, a NASA mission with international participation. 
The OGLE project has received funding from the National Science Centre, Poland, grant MAESTRO 2014/14/A/ST9/00121 to AU.
This work has made use of data from eROSITA, the primary instrument aboard SRG, a joint Russian-German science mission supported by the Russian Space Agency (Roskosmos), in the interests of the Russian Academy of Sciences represented by its Space Research Institute (IKI), and the Deutsches Zentrum für Luft- und Raumfahrt (DLR). The SRG spacecraft was built by Lavochkin Association (NPOL) and its subcontractors, and is operated by NPOL with support from the Max Planck Institute for Extraterrestrial Physics (MPE).
The development and construction of the eROSITA X-ray instrument was led by MPE, with contributions from the Dr. Karl Remeis Observatory Bamberg \& ECAP (FAU Erlangen-N{\"u}rnberg), the University of Hamburg Observatory, the Leibniz Institute for Astrophysics Potsdam (AIP), and the Institute for Astronomy and Astrophysics of the University of T{\"u}bingen, with the support of DLR and the Max Planck Society. The Argelander Institute for Astronomy of the University of Bonn and the Ludwig Maximilians University Munich also participated in the science preparation for eROSITA.
The eROSITA data shown here were processed using the eSASS software system developed by the German eROSITA consortium.
GV is supported by NASA Grant Number  80NSSC20K0803, in response to XMM-Newton AO-18 Guest Observer Program.
GV acknowledges support by NASA Grant number 80NSSC20K1107. 
TEW acknowledges support from the NRC-Canada Plaskett fellowship. MDS is supported by the Illinois Survey Science Fellowship of the Center for Astrophysical Surveys at the University of Illinois at Urbana-Champaign.
Software: XMM-Newton Science Analysis Software (SAS) v17, HEASoft v6.26, Stingray, Python v3.7.3, IDL\textsuperscript{\textregistered}

\section*{Data availability}

X-ray data are available through the High Energy Astrophysics Science Archive Research Center \url{heasarc.gsfc.nasa.gov}.
Other data underlying this article will be shared on reasonable request to the corresponding author.
The eROSITA data are subject to an embargo period of 24 months from the end of eRASS1 cycle. Once the embargo expires the data will be available upon reasonable request to the corresponding author.



\bibliographystyle{mnras}
\bibliography{SSS2}








\bsp	
\label{lastpage}
\end{document}